\documentclass[aps,pre,onecolumn,eps,graphicx]{revtex4}

\def\bec{\begin{center}}
\def\enc{\end{center}}

\def\ben{\begin{equation}}
\def\ba{\begin{array}}
\def\bea{\begin{eqnarray}}

\def\een{\end{equation}}
\def\eea{\end{eqnarray}}
\def\ea{\end{array}}
\def\btab{\begin{table}}
\def\btabu{\begin{tabular}}
\def\etab{\end{table}}
\def\etabu{\end{tabular}}
\def\bit{\begin{itemize}}
\def\eit{\end{itemize}}
\def\bef{\begin{figure}[htb]}
\def\befh{\begin{figure}[!h!]}
\def\enf{\end{figure}}

\def\b1{{\bf 1}}

\def\cos{\hbox{cos}\:}

\usepackage{epsfig,latexsym}

\newcommand \bew {\begin{widetext}}
\newcommand \enw {\end{widetext}}

\begin{document}

\title{\bf\noindent The field theoretic derivation of the
contact value theorem in planar geometries
and its modification by the Casimir effect}

\author{D.S. Dean$^{(1,2)}$ and R.R. Horgan$^{(2)}$}

\affiliation{(1) IRSAMC, Laboratoire de Physique Quantique, Universit\'e Paul Sabatier, 118 route de Narbonne, 31062 Toulouse Cedex 04, France\\
(2) DAMTP, CMS, University of Cambridge, Cambridge, CB3 0WA, UK \\
E-Mail:dean@irsamc.ups-tlse.fr, rrh@damtp.cam.ac.uk}

\date{22 September 2003}
\begin{abstract}
The contact value theorem for Coulomb gases in planar or film-like
geometries is derived using a Hamiltonian field
theoretic representation of the system. The case where the 
film  is enclosed by a material of different dielectric
constant to that of the film is shown to contain an additional 
Casimir-like term which is generated by
fluctuations of the electric potential about its mean-field
value.      
\end{abstract}  
\maketitle
\vspace{.2cm}
\pagenumbering{arabic}
\section{Introduction}
Coulomb gases arise in a huge variety of 
physical contexts from plasmas to soft condensed matter systems 
\cite{is,rusasc}.
Ideal Coulomb gases, where only point like charges are
considered, present a number of useful sum rules \cite{contact,deshol}. 
These sum rules are exact
identities between certain statistical mechanical observables. Sum
rules are useful for checking the validity of approximation schemes,
which must almost always be applied in the case  of 
strongly interacting systems. They also  provide potentially useful
experimental and numerical methods of indirectly measuring local 
observables in terms of macroscopically measurable quantities such as 
the pressure and vice-versa.
Finally sum rules may also be used to verify the accuracy of 
numerical simulations where, thanks to the sum rule,
the same quantity can be measured in two independent ways. 

An example  of such a sum rule is the contact value theorem which relates
the surface charge and density at the surface of the system
to the system's pressure \cite{contact,deshol}. 
In this paper we use a field theoretic approach to show how the basic 
contact value theorem can be derived in the  case of layered geometries, 
such as soap films, and show how the
contact value theorem is modified when the
dielectric constant ($\epsilon$) within
the film is different to that outside the film ($\epsilon_0$). The 
film geometry is of particular importance as it corresponds to the
experimental set up used to study soap films \cite{exps}. We  exploit the 
planar geometry to develop a Hamiltonian formulation of the Sine-Gordon
field theory which arises for Coulomb gases \cite{deho}. 
In this formulation the 
perpendicular direction, denoted here by $z$, acts as a temporal coordinate
in which a field $\phi(\bf r)$  on the plane perpendicular to 
$z$, and whose coordinates are denoted by the vector ${\bf r}$, propagates. 
The case
where global electro-neutrality holds  is treated within this
formulation. The condition of electro-neutrality  can be related to the 
choice of ground-state wave functional for the dynamical field 
$\phi({\bf r})$. 

In the canonical ensemble, the partition function for a system
of fixed particle number on a 3-d space ${\cal V}$ is given by \cite{deho}
\begin{equation}
Z = {\rm Tr}\int d[\phi] \exp\left(-{\beta\over 2} \int_{\cal V} d{\bf x}\  
\epsilon({\bf x})
\right(\nabla \phi({\bf x})\left)^2  + 
i\beta\int_{\cal V} d{\bf x}\ \rho_e({\bf
x})\phi({\bf x}) \right),\label{eqb}
\end{equation}
where $Tr$ denotes the classical trace over the particle positions
${1\over N!}\int_{\cal V}\prod_{i=1}^N d{\bf x}_i$ and $d[\phi]$ 
denotes the functional integral
over the field $\phi$. The field $\phi$ is the Wick rotated electrostatic field
$\psi$ and is thus related to $\psi$ via $\psi = -i\phi$.
The term $\rho_e({\bf x}) = \sum_{i=1}^N
q_i \delta({\bf x} - {\bf x}_i) + \rho_q({\bf x})$ is the charge density of
the system. The first term is the dynamical charge density 
which can vary in the system, ${\bf x}_i$ being the position of
particle $i$ and $q_i$ its charge. The second term $\rho_q$ is a 
quenched  background charge which is not dynamical and represents, for example,
a fixed surface charge. The integration volume  in the action $\cal V$
is all space. We note that the above treatment of a two component Coulomb
gas needs to be modified where the basic physical description of point charges
interacting via a Coulomb potential is thermodynamically unstable. For instance
the system can become unstable and have a tendency to collapse at low 
temperatures if some short range, for example hard core, repulsion is 
not included. The Sine-Gordon theory can always be regularized by introducing
a high momentum or short distance cut-off in the Fourier modes of the
field $\phi$. We note that the above formulation contains self interactions 
between the particles {\em i.e.} the terms $q_i q_j v({\bf r_i} - {\bf r_j})/2$
for $i=j$ where $v$ is the effective pairwise interaction are included. The 
interaction of a particle with its image charges is part of this contribution
and should be included, but the self-energy in the bulk medium 
should not contribute to the physical pressure and so must be subtracted.
For a monovalent system in dimension $d$ if one removes the bulk
self-energy term the Sine-Gordon free energy is corrected by a term
\begin{equation}
\Delta F = -N {e^2 v(0)\over 2} = -{Ne^2\over 2\epsilon} {S_d\over (2\pi)^d}
\int_0^\Lambda dk \ {k^{d-3}}\label{df}
\end{equation}
where $N$ is the number of particles, $S_d$ the surface of the unit sphere in
$d$ dimensions and $\Lambda$ is an ultra-violet or short distance cut-off.
Note that for $d>2$  $\Delta F$ is a regular function of $\Lambda$ and so
can be absorbed into the fugacity. However, for $d=2$ the integral in Eq. (\ref{df}) 
has an infra-red divergence at $k=0$ which must be cut-off at the inverse system 
size $1/L$ giving
\begin{equation}
\Delta F = -{Ne^2\over 4\pi\epsilon}\left({\ln(\Lambda) - \ln(1/L)}\right)\;.
\end{equation}
Thus, in two dimensions there is a correction to the Sine-Gordon pressure of 
\begin{equation}
\Delta P = -{\rho e^2\over 4\pi\epsilon} \label{dp2d}
\end{equation}
where $2\rho = N/L^2$ is twice the density of electrolyte..

Normally the particles $i$ are restricted to a sub-volume
$V$, for instance in the interior of a soap film for electrolyte solutions.
In the case where $\epsilon$ is constant, the functional integral 
over $\phi$ is easily done and we recover a system of $N$ particles
interacting via the Coulomb potential. For a varying dielectric
constant $\epsilon$  the resulting
interaction depends on the spatial variation of $\epsilon$ and the 
resulting integration gives rises to a more   complicated interaction
which can be interpreted in terms of image charges. The partition
function $Z$ in Eq. ({\ref{eqb}) thus contains a term which is due
to the pairwise inter-particle interaction, plus a functional determinant
coming from the integration over $\phi$. Both of these terms are present
in the physics of the problem and should be taken into account. The
form of Eq. ({\ref{eqb}) comes directly from a static approximation
to Quantum Electrodynamics (QED) where the non-zero frequency Matsubara
frequencies and electric currents (and thus the magnetic field) are neglected
in the action of QED \cite{deho,fehi,balo}. 
This approximation is justified when the 
charge distribution is very weakly be coupled to non-zero Matsubara
frequencies \cite{mani}, that is to say  the  response time of the charge
to the non-zero frequencies is  large. The non-zero  Matsubara
frequencies, when decoupled from the charge distribution yield a  
van der Waals interaction between the surfaces in the  problem and 
can be calculated independently in this approximation \cite{mani}. 
\section{Hamiltonian Field Theoretic Formalism}
For simplicity in what follows we consider a film system, that is to say an inner region having two interfaces separating it from  an outer region.  
The surface of the 
film has has area $A$ in the $(x,y)={\bf r}$ plane, the direction $z$ is
perpendicular to the film surface. 
For a system consisting of a film of thickness $L$, 
the film is in the region $z\in [0,L]$ and
the exterior of the film is in $[-T',0]$ (the left exterior) and
$[L,T-L]$ (right exterior). 
The total length of the physical system in the $z$ direction
$T+T'$  is taken to be constant and we consider the limit of $T$ and 
$T'$ large. In the simplest case, which we study here,
the electrolyte is monovalent and the fugacity of cations and 
anions is the same and denoted by $\mu$ within the film region
and is zero outside. This can be encoded in a spatially varying fugacity
$\mu({\bf x}) = \mu({\bf r},z) = \mu(z)$, with $\mu(z) = \mu$ 
for $z\in [0,L]$ and $\mu(z) = 0$ for $z\notin [0,L]$.
The dielectric constant within the 
film  is denoted by $\epsilon$ and
the external dielectric constant  is denoted by
$\epsilon_0$. For a soap  film, for example,  
$\epsilon_0$ could be the 
dielectric constant of air and $\epsilon$ the dielectric constant of
water. We may also consider systems where the regions $z\in [-\delta,0]$
and  $z\in [L,L+\delta]$ have a dielectric constant determined by the
dielectric constant of a surfactant and its concentration 
at the surface.

In the grand canonical ensemble the grand partition function is written
as 
\begin{equation}
\Xi = \int d[\phi] \exp\left(-S[\phi,L]\right).
\end{equation}
The film the pressure is therefore
given by
\begin{equation}
\beta P = {1\over A} {\partial\ln( \Xi)\over \partial L}
\end{equation}
In the case considered here,
$S$ is the action of a generalized Sine-Gordon field theory \cite{deho}
\begin{eqnarray}
S[\phi,L] &=& {\beta\over 2} \int_{\cal V} d{\bf x}\  \epsilon(z)
(\nabla \phi)^2 
-2  \int_{\cal V} d{\bf x} \ \mu(z) \cos\left(e\beta \phi({\bf x})\right)
\nonumber \\
&-& i\beta \int_{z=0} d{\bf r} \ \sigma \phi({\bf x})+
i\beta \int_{z=L} d{\bf r} \ \sigma \phi({\bf x}),
\end{eqnarray}
where  the integrations over 3-space, denoted by 
$d{\bf x}= d{\bf r}dz$,
are over the $(x,y)= {\bf r}$ plane of area $A$ and over the coordinate $z$.
The last terms are  the (constant uniform) 
surface charge contributions from  the 
transverse planes of area $A$ at $z=0$ and $z=L$.

We now rewrite the action using the $2+1$ decomposition in terms
of the field $\phi({\bf r})$ which evolves with a temporal coordinate $z$
\begin{equation}
\Xi = \int {\cal D}[\phi] \exp\left(-\int dz\  {\cal S}[\phi,z]\right),
\end{equation}
where the path integral action $\cal S$ is given by
\begin{eqnarray}
{\cal S}[\phi,z] &=& {\beta \epsilon(z)\over 2}
\int_A \left({\partial \phi\over 
\partial z}\right)^2 d{\bf r}
+ {\beta \epsilon(z)\over 2}\int_A d{\bf r}\ \left(\nabla_{\bf r}\phi\right)^2
- 2\mu(z) \int_A d{\bf r}\ \cos\left(e\beta \phi\right)\nonumber \\
&-& i\beta\sigma\left( \delta(z) + \delta(z-L)\right) \int_A
d{\bf r}\  \phi .
\end{eqnarray}   
The functional Schr\"odinger Hamiltonian for the path integral
outside the film is given by
\begin{equation}
H_E = \int_A d{\bf r} \left[-{1\over 2\beta \epsilon_0}
{\delta^2 \over \delta \phi({\bf r})^2}
+ {\beta \epsilon_0 \over 2}\left(\nabla_r \phi\right)^2\right].
\end{equation}
Inside the film the Hamiltonian is
\begin{equation}
H_F = \int_A d{\bf r} \left[-{1\over 2\beta \epsilon}
{\delta^2 \over \delta \phi({\bf r})^2}
+ {\beta \epsilon \over 2}\left(\nabla_r \phi\right)^2
- 2\mu\cos\left(e\beta \phi\right)\right]
\label{eqhf}.
\end{equation}
The terms containing  the surface charge and other more general surface
interactions may be included as source terms. We note here that in
the case of the 1-d
Coulomb gas the field $\phi$ is interpreted as  the position of a 
single particle and
the quantum mechanical formalism then leads to an exact solution
\cite{oned}.
In this formulation we postulate that
\begin{equation}
\Xi = {\rm Tr}\exp\left(-TH_E\right){\cal O}\exp\left(-LH_F)\right)
{\cal O}\exp\left(-(T'-L)H_E\right) \label{eqbasic},
\end{equation}
where here ${\rm Tr}$ denotes the trace over a complete set of states.
The source term ${\cal O}$ for a constant surface charge is clearly given by
\begin{equation}
{\cal O} = \exp\left(i\beta\sigma\int_A d{\bf r}\ \phi({\bf r}).
\right)
\end{equation}  
The reason we say that Eq. (\ref{eqbasic}) is postulated, is that the
boundary conditions for the above path integral are not straightforward
to determine. We shall see later that the above choice assures the 
electro-neutrality of the film since in the limit $T,T'\to\infty$ 
it only involves the
ground state wave functional $|\Psi_0 \rangle$ of the Hamiltonian $H_E$. 
In general, instead of taking the trace, 
we  could specify any linear combination of wave functionals of 
$H_E$ as the  initial (and final) state 
for the field $\phi$, however as long as 
it has a non-zero component of the wave function  $|\Psi_0 \rangle$, in the
limit where $T$ and $T'$ are large the result will be the same.
So, in the limit of large $T$ and $T'$  the  grand partition function 
may thus be written as
\begin{equation}
\Xi = \langle \Psi_0| \exp\left(-TH_E\right){\cal O}\exp\left(-LH_F\right){\cal O}\exp\left(-(T'-L)H_E\right)
|\Psi_0 \rangle. .
\end{equation}
In this formulation, the derivative with respect to $L$ is easily and
unambiguously taken, we find that
\begin{eqnarray}
{\partial \Xi\over \partial L}
&=&-\langle \Psi_0| \exp\left(-TH_E\right){\cal O}\exp\left(-LH_F)\right)
H_F{\cal O}\exp\left(-(T'-L)H_E\right)
|\Psi_0 \rangle 
\nonumber \\&+& \langle \Psi_0| \exp\left(-TH_E\right){\cal O}\exp\left(-LH_F\right){\cal O}
H_E\exp\left(-(T'-L)H_E\right)
|\Psi_0 \rangle .\label{eqdxi1}
\end{eqnarray}
We now define the momentum operator for the field $\phi$ at the point ${\bf r}$
by
\begin{equation}
P_{\phi({\bf r})} = -i {\delta\over \delta \phi({\bf r})}.
\end{equation}
This leads to  the commutation relation
\begin{equation}
[P_{\phi({\bf r})}, \phi({\bf r'})] = -i\delta({\bf r} - {\bf r'}).
\end{equation}
The kinetic operator $K$ is then defined by
\begin{equation}
K = \int_A d{\bf r} \ P^2_{\phi({\bf r})} = -\int_A d{\bf r}\ {\delta^2\over \delta
\phi({\bf r})^2}.
\end{equation}
In this notation the exterior and interior Hamiltonians read
\begin{eqnarray}
H_E &=& {K\over 2\beta \epsilon_0} + V_E \\
H_F &=& {K\over 2\beta \epsilon} + V_F,
\end{eqnarray}
where the functional potentials of the Hamiltonians are given by
\begin{eqnarray}
V_E &=& {\beta \epsilon_0\over 2} \int_A d{\bf r}\  (\nabla_{\bf r} \phi)^2
 \\
V_F &=& {\beta \epsilon\over 2} \int_A d{\bf r} \ (\nabla_{\bf r} \phi)^2
-2\mu \int_A d{\bf r}\ \cos\left(e\beta \phi({\bf r})\right).
\end{eqnarray}
We note that these functional potentials are pure functionals and involve
no functional derivative operators, thus they commute with other functionals,
notably ${\cal O}$.
The Eq. (\ref{eqdxi1}) is now written as
\begin{equation}
{\partial \Xi\over \partial L}= -\langle \Psi_0| 
\exp\left(-TH_E\right){\cal O}\exp\left(-LH_F\right)\left([H_F,{\cal O}]
-{\cal O}(H_E-H_F)\right)\exp\left(-(T'-L)H_E\right)
|\Psi_0 \rangle .
\end{equation}
Using the result
\begin{equation}
[ P^2_{\phi({\bf r})},{\cal O}] = 2\beta\sigma{\cal O}  P_{\phi({\bf r})}
+ \beta^2\sigma^2 {\cal O},
\end{equation}
we obtain
\begin{eqnarray}
{\partial \Xi\over \partial L} &=& -A {\beta \sigma^2 \over 2  \epsilon}
\Xi - {\sigma\over \epsilon}\langle \Psi_0| 
\exp\left(-TH_E\right){\cal O}\exp\left(-LH_F\right){\cal O} 
\left(\int_A d{\bf r}
P_{\phi({\bf r})}\right) \exp\left(-(T'-L)H_E\right)
|\Psi_0 \rangle \nonumber \\
&+& \langle \Psi_0| 
\exp\left(-TH_E\right){\cal O}\exp\left(-LH_F\right){\cal O} 
\left(H_E -H_F\right) \exp\left(-(T'-L)H_E\right)
|\Psi_0 \rangle. \label{eqdxi2}
\end{eqnarray}
The final result for the pressure is thus
\begin{equation}
\beta P = - {\beta \sigma^2\over 2 \epsilon} + {1\over A}\langle (H_E -H_F)
\vert_{s^+}\rangle
- {1\over A}{\sigma\over  \epsilon}\langle \int_A d{\bf r} \ 
P_{\phi({\bf r})}\vert_{s^+} \rangle, 
\end{equation}
where the above notation indicates the normalized expectation values of the 
operators shown, evaluated at the rightmost outer-surface of the film $s^+$
{\em i.e.} at $z= L^+$.
The third term above can be shown to be zero in the case of systems which are
globally electro-neutral; we will demonstrate this more technical point later.
We note the relation
\begin{equation}
H_F = {\epsilon_0\over \epsilon} H_E - 
\int_A d{\bf r}\ \left[ 2\mu \cos\left(e\beta \phi\right) -{\beta \epsilon
\over 2}\left(1- {\epsilon_0^2\over \epsilon^2}\right)\left(\nabla_{\bf r}\phi
\right)^2\right].
\end{equation}
Also it is straightforward to see that
\begin{equation}
\langle \mu \exp(\pm i e \beta \phi({\bf r})\vert_z \rangle = 
\langle \rho^\pm({\bf r},z) \rangle,
\end{equation}   
is the mean  value of the cation/anion density at the point $({\bf r},z)$.
Putting these results together we obtain
\begin{equation}
\beta P = \langle \rho^+\vert_s + \rho^-\vert_s\rangle - 
{\beta \sigma^2\over 2\epsilon} 
- {\beta\epsilon\over 2} 
\left(1 - {\epsilon_0^2\over \epsilon^2}\right)\langle (\nabla_{\bf r} \phi)^2\vert_{s} \rangle
+ {1\over A} 
\left(1-{\epsilon_0\over \epsilon}\right)\langle H_E\vert_{s^+}
\rangle,
\label{contact1}
\end{equation}
where all but the last term  are evaluated on
the  surface $s$ ($z=L$) at any given point (by homogeneity in the plane $A$).
Here there is no ambiguity with regards to the interior or exterior of the
surface as the terms are pure functionals of the field
$\phi$ and  commute with ${\cal O}$ as it is, itself, a pure functional 
of the field $\phi$. The last term is evaluated at the outer-surface 
$s^+$ (at $z=L^+$). We thus obtain
\begin{equation}
\beta P = \langle \rho^+\vert_s + \rho^-\vert_s\rangle - 
{\beta \sigma^2\over 2\epsilon} 
- {\beta\epsilon\over 2} 
\left(1 - {\epsilon_0^2\over \epsilon^2}\right)\langle (\nabla_{\bf r} \phi)^2\vert_{s} \rangle
+{E_0\over A}\left(1-{\epsilon_0\over \epsilon}\right),
\label{contact2}
\end{equation}
where $E_0$ is the energy of the ground-state wave functional $|\Psi_0\rangle$.
In the case where $\epsilon_0=\epsilon$, Eq. (\ref{contact2}) immediately
yields the classic contact value theorem \cite{contact} 
as the third and fourth terms are identically 
zero. 

While the functional terms appearing in the expectation values of the 
Hamiltonians are easy to interpret in terms of observables,
the kinetic term $K$ requires more work. The key result here is 
\begin{equation}
\langle {1\over 2\beta \epsilon(z)}  P^2_{\phi({\bf r})}
\vert_z \rangle =- \langle {\beta \epsilon(z)\over 2} \left({\partial 
\phi({\bf r},z)
\over \partial z}\right)^2\rangle. \label{eqfd}
\end{equation}  
This can be seen in the Heisenberg formalism which gives
\begin{equation}
\langle\left({\partial \phi({\bf r},z)
\over \partial z}\right)^2\rangle = \langle [H,\phi({\bf r})]^2\vert_z
\rangle.
\end{equation}

We therefore have that for a point $z$ inside the film
\begin{equation}
\langle H_F\vert_z\rangle = \langle  \int_A d{\bf r}\left[ -{\beta \epsilon\over 2} \left({\partial \phi({\bf r},z) \over \partial z}\right)^2
+{\beta \epsilon\over 2}\left(\nabla_{\bf r} \phi\right)^2 - 2\mu \cos(
e\beta \phi)\right]\vert_z \rangle,
\label{hfe}
\end{equation}
and outside the film
\begin{equation}
\langle H_E\vert_z\rangle = \langle \int_A d{\bf r}\left[ -{\beta 
\epsilon_0\over 2} \left({\partial \phi({\bf r},z) \over \partial z}\right)^2
+{\beta \epsilon_0\over 2}\left(\nabla_{\bf r} \phi\right)^2 \right]\vert_z
\rangle.\label{hfe2}
\end{equation}

Using Eq. (\ref{hfe2})  in Eq. (\ref{contact1}) we find the alternative 
expression
\begin{equation}
\beta P = \langle \rho^+\vert_s + \rho^-\vert_s\rangle - 
{\beta \sigma^2\over 2\epsilon} 
- {\beta\over 2} 
\left(\epsilon-\epsilon_0\right)
\langle (\nabla_{\bf r} \phi)^2\vert_{s} \rangle
-{\beta \epsilon_0\over 2}\left(1-{\epsilon_0\over \epsilon}\right)
\langle \left({\partial \phi\over \partial z}\right)^2\vert_{s^+}\rangle.
\label{contact3}
\end{equation}

In mean field theory, when the field $\phi$ is replaced by its mean field
electrostatic field $-i\psi_{c}$, the electro-neutrality of the system 
implies that $\partial \psi_{c}/\partial z = 0$ outside the film and 
the homogeneous nature of the mean field solution in the plane of the 
film yields $\nabla_{\bf r}\psi_{c} = 0$. Thus the third and fourth terms
of Eq. 
(\ref{contact3}) are zero and hence  the contact value theorem
as classically stated is always verified at the mean-field level. If we 
expand about the mean field solution, we see that the correction term
due to the variation of the dielectric constants comes from fluctuations about
the mean-field solution. The value of this term can be calculated 
in the case of weak electrolyte strength in the Debye-H\"uckel
approximation \cite{mani,dh,deho}.

We now return to the question of the global electro-neutrality of the
system. Using the correspondence between the field 
$\phi$ and $\psi$ we have
\begin{equation}
\langle {\partial \psi({\bf r},z)\over \partial z}\rangle =
-i \langle {\partial \phi({\bf r},z)\over \partial z}\rangle.
\end{equation}
If the film is electro-neutral then, by symmetry about the film's mid-plane,
the integral of the electric field over every plane perpendicular
to the $z$ direction outside the film must vanish, and thus
\begin{equation} 
\langle \int_A d{\bf r}\ 
{\partial \psi({\bf r})\over \partial z}\vert_z\rangle = 0,
\end{equation}
for $z > L$ (and also for $z < 0$). 
Using the Heisenberg formalism we find
\begin{equation} 
-i\langle \int_A d{\bf r}\ {\partial \phi({\bf r})\over \partial z}
\vert_z\rangle = i 
\langle\int_A d{\bf r} \ [H_E, \phi({\bf r})]\vert_z\rangle
= {1\over \epsilon_0 \beta}\langle \int_A d{\bf r} P({\bf r})\vert_z \rangle,
\end{equation}
and using  the fact that
\begin{equation}
\exp\left(-(T'-L)H_E\right)|\Psi_0 \rangle =   
\exp\left(-(T'-L)E_0\right)|\Psi_0 \rangle,
\end{equation}
we may write
\begin{equation}
\int_A d{\bf r} P_{\phi({\bf r})}\exp\left(-(T'-L)H_E\right)|\Psi_0 \rangle =
\exp\left(-(T'-L)H_E\right) \int_A d{\bf r} P_{\phi({\bf r})}|\Psi_0 \rangle.
\end{equation}
The final step in the proof of our result is to show that
\begin{equation}
\int_A d{\bf r}\ P_{\phi({\bf r})}|\Psi_0 \rangle = 0 \label{eqen}.
\end{equation}
Outside the film we consider the Fourier representation of the 
field 
\begin{equation}
\phi({\bf r}) = {1\over \sqrt{A}}\sum_{{\bf p}} \tilde{\phi}({\bf p})
\exp(i{\bf p}\cdot {\bf r}).
\end{equation}
In this basis the exterior Hamiltonian becomes
\begin{equation}
H_E = \sum_{\bf p}\left[-{1\over 2\beta\epsilon_0}{\delta^2\over \delta
 \tilde{\phi}({\bf p})\delta\tilde{\phi}(-{\bf p})} + 
{ \beta \epsilon_0
{\bf p}^2 \over 2}\tilde{\phi}({\bf p})\tilde{\phi}(-{\bf p})\right]
\end{equation}
Thus outside the film  each  Fourier mode of $\phi$ is the coordinate of an
independent Harmonic
oscillator and the ground state wave functional
is given by
\begin{equation}
\langle {\tilde \phi}|\Psi_0\rangle \propto \prod_{\bf p} \exp\left(-{1\over 2}M\omega({\bf p})\tilde{\phi}({\bf p})\tilde{\phi}(-{\bf p})\right), \label{eqgs}
\end{equation}
where $M = \beta\epsilon_0$ and $\omega({\bf p}) = |{\bf p}|$.
We now note that
\begin{equation}
\int_A d{\bf r} P_{\phi({\bf r})} = -i\sqrt{A} {\delta\over \delta\tilde{\phi}({\bf 0})}.
\end{equation}
However  from Eq. (\ref{eqgs}) $|\Psi_0 \rangle$ is
clearly independent of $\tilde{\phi}({\bf 0})$  and so the desired  
result Eq. (\ref{eqen}) follows directly.

Using the fact that the Hamiltonian $H_F$ commutes with itself we may also
rewrite Eq. (\ref{eqdxi1}) as
\begin{eqnarray}
{\partial \Xi\over \partial L}
&=&-\langle \Psi_0| \exp\left(-TH_E\right){\cal O}
\exp\left(-{1\over 2}LH_F)\right)H_F\exp\left(-{1\over 2}LH_F)\right){\cal O}
\exp\left(-(T'-L)H_E\right)
|\Psi_0 \rangle 
\nonumber \\&+& \langle \Psi_0| \exp\left(-TH_E\right){\cal O}\exp\left(-LH_F\right){\cal O}
H_E\exp\left(-(T'-L)H_E\right)
|\Psi_0 \rangle. \label{eqmp1}
\end{eqnarray} 
Applying the results derived above and using Eq. (\ref{eqhf}) leads to
\begin{equation}
\beta P = \langle \rho_+|_m + \rho_-|_m \rangle - {\beta\epsilon\over 2}
\langle \left(\nabla_{\bf r} \phi\right)^2|_m \rangle +  {\beta\epsilon\over 2}
\langle \left({\partial \phi\over \partial z}\right)^2|_m \rangle
+{\beta\epsilon_0\over 2}
\langle \left(\nabla_{\bf r} \phi\right)^2|_{s} \rangle 
-  {\beta\epsilon_0\over 2}
\langle \left({\partial \phi\over \partial z}\right)^2|_{s^+} \rangle,
\label{eqmpf}
\end{equation}
where $m$ in the above indicates the value taken at the mid-plane of
the film $z=L/2$. At the mean field level for an electro-neutral film
the last term again vanishes and, in addition, by symmetry the mean field
solution $\psi_c$ obeys $\partial\psi_c/ \partial z = 0$ at $z=L/2$. Hence if
we neglect the corrections to mean field in the second two terms, the 
so called mid-plane formula for the mean-field approximation \cite{is,rusasc}
is immediately recovered. It is also worth noting that Eq. (\ref{eqmpf}) is
valid for any surface charge operator $\cal{O}$, thus explaining why various
charge regulated models obey the classic mid-plane formula at the mean-field
level \cite{vsc,deho}.

The Hamiltonian formalism is also illuminating in the case of a bulk 
calculation. Here the grand partition function is given by
\begin{equation}
\Xi_B = {\rm Tr} \exp(-LH_F)
\end{equation}
which gives the pressure bulk pressure as
\begin{equation}
\beta P_B = -{1\over A} \langle H_F \rangle
\end{equation}
where the expectation above is taken at any value of $z$. Now using the 
relation Eq. (\ref{hfe}) and the homogeneity in the transverse plane $A$ 
we find
\begin{equation}
\beta P_B = 2 \rho -{\beta \epsilon\over 2}\langle \left(\nabla_{\bf r}\phi
\right)^2\rangle + {\beta \epsilon\over 2}   \langle \left( \partial \phi\over \partial z\right)^2 \rangle
\end{equation}
where $\rho$ is the bulk electrolyte density.
Using the isotropy of the bulk system we finally obtain
\begin{equation}
\beta P_B = 2 \rho - 
{\beta \epsilon\over 6}\langle \left(\nabla \phi
\right)^2\rangle,\label{pd3}
\end{equation}
where the gradient  above is the full three dimensional gradient. The first 
term above can be interpreted as an osmotic term and the second term has its
origin in the Maxwell stress tensor for the electrostatic field \cite{lali}.

In $d$ dimensions, using exactly the same decomposition in terms of
a temporal coordinate $z$ and a $d-1$ dimensional hyper-surface, 
the above expression becomes
\begin{equation}
\beta P_B = 2 \rho - 
{\beta \epsilon\over 2d}(d-2)\langle \left(\nabla \phi
\right)^2\rangle.
\label{pbd}
\end{equation}
Note that the above expression recovers \cite{2d} the physical pressure for
a two-dimensional neutral Coulomb gas which is given by the simple formula
\begin{equation}
P_p = 2\rho k_BT(1  -  {e^2\over 8\pi\epsilon k_BT })
\end{equation} 
where the Sine-Gordon pressure has been corrected by the 2 dimensional 
self energy term of Eq. (\ref{dp2d}). 
This result can be simply understood from the logarithmic nature
of the Coulomb potential in two dimensions and is valid in the 
region of the thermodynamic stability of a purely Coulomb system in two 
dimensions \cite{2d}. 

The expression Eq. (\ref{pd3}) may also be used to calculate the physical bulk 
pressure of a 3-dimensional Coulomb gas within the Debye-H\"uckel 
approximation. One finds that
\begin{equation}
\beta P_B = 2\rho -{1\over 12 \pi^2}\int_0^\Lambda dk\ {k^4 \over k^2 + m^2}
\end{equation}
where $\Lambda$ is again an ultra-violet cut-off and $m = \sqrt{2 \rho e^2 \beta/\epsilon}$ is the Debye mass. This gives for large $\Lambda$
\begin{equation}
\beta P_B = 2\rho -{1\over 12 \pi^2}\left({\Lambda^3\over 3} - m^2 \Lambda
+ m^3{\pi\over 2}\right)
\end{equation}
which is clearly divergent. However the physical pressure due to the ions
$P_p$ is given by
\begin{equation}
\beta P_p = \beta P_B - \beta P_B(0) - \beta P_S
\end{equation}
where $\beta P_B(0)$ is the pressure due to  an electrostatic field 
in the absence
of ions and $P_S$ is a contribution coming from the self interaction of the
ions with themselves. To see the origin of $P_S$ we note that
\begin{equation}
\epsilon \nabla \phi = -i\epsilon \nabla \psi = i {\bf E}
\end{equation}
where ${\bf E}$ is the electric field. Clearly ${\bf E} = \sum_i {\bf E}_i$
where ${\bf E}_i$ is the field due to the  single particle $i$. The term in 
${\bf E}^2$ which contributes to the interaction between particles is thus
${\bf E}^2 - \sum_i {\bf E}_i^2$. This means that 
\begin{equation}
P_S = {1\over 6\epsilon}\langle \sum_i {\bf E}_i^2\rangle =   
{N\over 6\epsilon} \langle {\bf E}_i^2\rangle
\end{equation}
Hence if $\psi'$ is the electrostatic potential generated by a single particle
at the position of interest then
\begin{eqnarray}
\beta P_S &=& {2\rho\epsilon\beta\over 6}\int d{\bf r} \ (\nabla \psi')^2 \nonumber \\
&=&  {\rho e^2\beta \over 6 \pi^2 \epsilon}\int_0^\Lambda dk = {1\over 12 \pi^2 } m^2 \Lambda
\end{eqnarray} 
Putting all this together gives us the well known Debye pressure formula
\begin{equation}
\beta P_p = 2\rho - {m^3\over 24\pi}
\end{equation}
     
We remark here that the general result
Eq. (\ref{pbd}) may also be derived by putting the field theory on a lattice
and changing the volume of the system by varying the lattice size \cite{dehop}.
\section{Conclusion}

We have shown  in the case of planar geometries that the 
Sine-Gordon type field theory for Coulomb systems can be expressed
in terms of a $2+1$ path integral for a field $\phi$ on the 2-d surface
parallel to the film surface. The expression has the form of an S-matrix
and the choice of the ground-state wave functional for the ingoing and
outgoing states of the field $\phi$ imposes the global electro-neutrality
of the system. In the case of constant dielectric constant throughout
the system, the classic contact value theorem is recovered. In the 
presence of  dielectric variations, we see how the contact value theorem
is corrected by fluctuations about the mean-field solution. A corresponding
version of the mid-plane formula is also derived, 
which is also shown to be generically valid at the 
mean-field level. We again see that the corrections come from fluctuations
about the mean-field solution. In both cases these fluctuations 
about the mean-field solution are known to be of Casimir type and can be
calculated within various approximations. The results show clearly that
the pressure of the system has two distinct, 
though inter-related, contributions:
an osmotic pressure term plus a term coming from static thermal 
fluctuations of the field $\phi$. The present Hamiltonian technique has 
the advantage of giving an unambiguous way of taking the derivative
of the grand partition function with respect to the film thickness $L$
in the presence of surface charges and dielectric discontinuities. In
the usual field theoretic formulation the taking of this derivative and
the interpretation of the resulting terms as thermodynamic averages is
far from obvious. In this Hamiltonian approach, the passage 
between the contact value result and mid-plane result is also
straightforward. We have also shown how the Hamiltonian approach provides
an alternative method for representing the pressure of bulk systems, giving
results in accordance with those obtained via other methods,

As far as future work is concerned, using the results obtained here similar
results can be obtained for more complicated layer geometries (additional
dielectric layers for instance) and also for more complicated surface 
charges, for instance modulated surface charges and surface charges
built up from thermodynamic or chemical surface charging mechanisms
({\em i.e.} charge regulated models) \cite{vsc,deho}.

\pagestyle{plain}

\end{document}